\documentclass[rmp, aps, nofootinbib,11pt, a4paper]{revtex4}
\usepackage{epsfig}
\usepackage{amsmath,amssymb}
\usepackage{fancyhdr}
\usepackage[normalem]{ulem}
\usepackage{color}
\usepackage{mathrsfs}  
\usepackage{hanging}
\usepackage[colorlinks=true, linkcolor=blue, citecolor=magenta]{hyperref}
\usepackage{url}

\usepackage{pifont}

\usepackage{float, placeins}
\usepackage{mwe}
\usepackage[caption=false]{subfig}
\usepackage{morefloats}
\usepackage{empheq}
\usepackage{bm}
\usepackage{booktabs} 
\usepackage{multirow}
\usepackage{makecell}
\usepackage{cases}
\usepackage[normalem]{ulem}
\usepackage[margin= 1in]{geometry}
\usepackage{setspace}

\makeatletter
\gdef\@ptsize{1}
\makeatother

\allowdisplaybreaks

\begin{document}

\title{{\large Spacetime singularities and a novel formulation of indeterminism}}

\author{Feraz Azhar}
\email[Email address: {fazhar@nd.edu}]{}\affiliation{Department of Philosophy, University of Notre Dame, Notre Dame, IN, 46556, USA \&\\ Black Hole Initiative, Harvard University, Cambridge, MA, 02138, USA}
\author{Mohammad Hossein Namjoo}
\email[Email address: {mh.namjoo@ipm.ir}]{}\affiliation{School of Astronomy, Institute for Research in Fundamental Sciences (IPM), Tehran, Iran}

\date{\today}
\begin{abstract}
\vspace{-1cm}
\singlespacing
Spacetime singularities in general relativity are commonly thought to be problematic, in that they signal a breakdown in the theory. We address the question of how to interpret this breakdown, restricting our attention to classical considerations (though our work has ramifications for more general classical metric theories of gravity, as well). In particular, we argue for a new claim: spacetime singularities in general relativity signal {\it indeterminism}.

\hspace{2ex} The usual manner in which indeterminism is formulated for physical theories can be traced back to Laplace. This formulation is based on the {\it non-uniqueness} of future (or past) states of a physical system---as understood in the context of a physical theory---as a result of the specification of an antecedent state (or, respectively, of a subsequent state). We contend that for physical theories generally, this formulation does not comprehensively capture the relevant sense of {\it a lack of determination}. And, in particular, it does not comprehensively capture the sense(s) in which a lack of determination (and so, indeterminism) arises due to spacetime singularities in general relativity.

\hspace{2ex}We thus present a novel, broader formulation, in which indeterminism in the context of some physical theory arises whenever one of the three following conditions holds: future (and/or past) states are  (i) not unique---as for Laplacian notions of indeterminism; (ii) not specified; or (iii) `incoherent'---that is, they fail to satisfy certain desiderata that are  internal to the theory and/or imposed from outside the theory. We apply this formulation to salient features of singularities in general relativity and show that this broader sense of indeterminism can comprehensively account for (our interpretation of) the breakdown signaled by their occurrence.
\end{abstract}

\maketitle
\vspace{-1.7cm}
\singlespacing
\tableofcontents
\section{Introduction}\label{SEC:Intro}

Singularities in general relativity (GR)---and in metric theories of gravity more broadly---remain largely mysterious and are commonly thought to signal the breakdown of such theories.\footnote{Metric theories of gravity, as described by~\citet{will_14}, are those that satisfy three criteria: (i) there exists a symmetric metric tensor field on a manifold; (ii) test particles follow geodesics of this metric field; (iii) in local Lorentz frames, the (non-gravitational) laws of physics employ special relativity. GR is, of course, one such metric theory, as are, for example, effective-field-theoretic generalizations of GR. In this paper we will focus primarily on GR.} Two notable circumstances in which they are thought to arise are in the final stages of the evolution of a sufficiently massive star, as for black holes, and in the very earliest moments of our universe, in the form of a `big-bang singularity'. In the former case, the process of gravitational collapse---a process that we believe has occurred many times in our universe---yields a class of objects that are not wholly described by the theory that posits their existence (namely, by GR). A similar claim holds in the latter case---in which the description of spacetime as a whole, provided by GR, appears to break down. A notable feature of such circumstances is that they are not necessarily a consequence of special, unrealistic circumstances, a claim encapsulated in a set of remarkable theorems proved and elaborated on in the five years from 1965 to 1970---mainly by Penrose, Hawking, and Geroch---culminating in the Hawking-Penrose theorem~\cite{hawking+penrose_70}. And so, in analyzing such ubiquitous and mysterious entities, a foundational question arises: in what sense does GR break down as a result of singularities?

A new response to this question, which we develop in this paper, is that singularities signal {\it indeterminism} in GR. This is a thorny task, for the doctrine of determinism (and that of indeterminism---which we will take to be its negation\footnote{Given this relationship between determinism and indeterminism, in this paper we will freely refer to one or the other concept as is appropriate in the relevant context.}) has a long and controversial history. Such controversy continues to manifest in our understanding of the two pillars of twentieth century physics, namely, GR and quantum mechanics.  [See~\citet[Ch.~2]{salmon_98} for a broad overview of determinism in science, and~\citet{earman_86} for a more detailed treatise.] Here, we will focus exclusively on classical considerations, though within this domain of applicability we expect our remarks will apply relatively broadly (namely, to a broad class of metric theories of gravity).

The standard account of determinism in physics---and, indeed, the one with respect to which aspects of GR are usually analyzed---can be traced back to~\citet{laplace_1814}. This notion of `Laplacian determinism', as it applies to {\it theories} (as opposed to, say, the world itself), amounts to the following idea: a deterministic theory that describes some physical system is one in which, given a state of the physical system at some moment in time, the future (and possibly the past) states of the physical system are {\it uniquely prescribed}. Of course, much remains unclear in this description, for we need to make precise what we mean by a `state' of the `physical system', as well as a `moment in time': for now, our intuitive notions of such concepts will suffice. (We will elaborate on these concepts in Sec.~\ref{SEC:LD+}.) In this paper we will argue that Laplacian determinism and recent closely related formulations [as in, for example,~\citet{butterfield_89} and~\citet{doboszewski_19}] are limited in their ability to describe salient features of physical theories that signal {\it a lack of determination} (on the part of the theory). In particular, we will argue that such formulations do not comprehensively capture the senses in which a lack of determination (and so, indeterminism) arises as a result of spacetime singularities in GR.\footnote{Note that neither~\citet{butterfield_89} nor~\citet{doboszewski_19} necessarily endorse their descriptions of determinism as those that are applicable to discussions about singularities; the more substantive claim in the main text, which will again be touched upon in Sec.~\ref{SEC:LD+}, is that their descriptions are closely related to a sense of Laplacian determinism.} 

There are thus two parts to our project. The first is a description of how we understand indeterminism. The second is an application of this formulation of indeterminism to salient features of singularities in GR. With regard to the first part, we will argue that the usual Laplacian notion of determinism does not account for a more comprehensive sense in which one can understand the relevant notion of a `determination'.\footnote{An analogous problem arises in converting our intuitions about singularities in GR into a precise definition of a singularity. [See, for example,~\citet{hawking+ellis_73},~\citet{wald_84}, and~\citet{earman_95}.]} Our emphasis accords with the spirit of the following claim by Earman:
\begin{quote}
Laplacian determinism and its close relatives are, to my knowledge, the only varieties which have received attention in the philosophical literature. The explanation cannot be that no other variety is relevant to the analysis of modern science~\dots ~\cite[p.~17]{earman_86}.
\end{quote}
The formulation of determinism we describe is broader than the Laplacian notion and (roughly) corresponds to a disjunction of the uniqueness, existence, and what we term the `coherence' of a theory's specification of states. (Indeterminism then arises whenever at least one of these disjuncts is not satisfied.) With regard to the second part of our project, we will enumerate salient features of singularities in GR and show how the above sources of indeterminism arise for each of the identified features. We thus aim to establish that there is, indeed, a sense in which singularities in GR signal a type of indeterminism and that this type of indeterminism cannot solely be understood as some variety of Laplacian indeterminism.

To this end, in Sec.~\ref{SEC:Singularities}, we describe how singularities are usually understood in GR, and distinguish two classes of properties that apply to them: namely, geometrical properties and `causal' properties. In Sec.~\ref{SEC:Indeterminism}, we describe and develop our formulation of indeterminism for physical theories, which is broader than the usual Laplacian account. In Sec.~\ref{SEC:Sing}, we provide a  description of salient features of singularities in GR, which, we argue, signal a failure of determinism---as understood in the context of the formulation of (in)determinism developed in Sec.~\ref{SEC:Indeterminism}.  In Sec.~\ref{SEC:Varying}, we describe how certain physical processes in GR can change sources of indeterminism---in the case of certain black-hole solutions and for Cauchy horizons. We conclude with a brief summary of our overall argument in Sec.~\ref{SEC:Conclusion}.

\section{Singularities in general relativity}\label{SEC:Singularities}

Singularities are arguably general relativity's most problematic feature and they remain over a century's old discontent of the cognoscenti. Perhaps the first more comprehensive statement about singularities for metric theories of gravity---and in particular for GR---was the Hawking-Penrose theorem~\citep{hawking+penrose_70}. This theorem provides conditions under which singularities arise, showing that they are not a feature of special initial conditions. Strikingly, the theorem does not provide detailed information about the nature of the singularities that arise: just the conditions that lead to singularities, which are diagnosed via curves that `stop short' (in particular, via timelike or null geodesics). (We will say more about precisely what it means for a curve to stop short in the following subsection.) This lack of a specification of the nature of the singularity is, indeed, a general feature of singularity theorems in GR (including those that have been  developed since the theorem by Hawking and Penrose). Questions thus remain as to the relationship between the manner in which singularities in these theorems are diagnosed and more detailed information about the nature of spacetime at or near singularities [see, for example,~\citet{senovilla_98, senovilla_12}]. 

In this section, to prepare for what follows, we will distinguish two broad classes of properties that can be ascribed to singularities in GR: (i) geometrical properties and (ii) what we term `causal' properties.

\subsection{Geometrical properties of singularities}\label{SEC:SingularitiesGEO}

Singularities as understood from a geometric perspective---and not just those that arise via gravitational collapse---are perhaps (still) most comprehensively classified, in the context of GR, by~\citet{ellis+schmidt_77}. In their classification scheme, for maximally extended spacetimes (roughly, spacetimes that cannot be embedded into a larger spacetime)\footnote{The focus on maximally extended spacetimes is to avoid classing as singular, an entity or region that would be described as nonsingular if the spacetime could be extended, yielding a `larger' one. For example, Minkowski spacetime with a single point removed is not maximally extended and so does not count as singular. For maximally extended spacetimes, singularities that arise are thus intrinsic to the spacetime.}, singularities fall into one of two classes: 
\begin{itemize}
\item[(i)] quasi-regular singularities---the mildest type of singularity, where certain components of the Riemann tensor are well-behaved\footnote{\label{FN:Riemm}We refer, here, to components of the Riemann tensor in an orthonormal frame that is parallely propagated along any curve that ends at the singularity [see~\citet{ellis+schmidt_77} for further details].}---such as one encountered at the tip of a cone; or 
\item[(ii)] curvature singularities, where there are pathologies related to the components of the Riemann tensor mentioned in (i).
\end{itemize}
This latter category is composed of either non-scalar singularities (where curvature scalars are not pathological) or scalar singularities (where curvature scalars are pathological). We will say more about the sense in which such pathologies may arise in Sec.~\ref{SEC:PhysQuant}, but a common way in which such pathologies manifest is via a divergence of such scalars, as one approaches the singular region.

In this classification scheme, all such singularities are diagnosed via incomplete curves, namely, curves that `stop short'. Broadly, such a curve cannot be extended to infinitely large values of the parameter that distinguishes different points along the curve. Precisely which types of curves are used to diagnose singularities in this way is somewhat contentious. The classification scheme of~\citet{ellis+schmidt_77} includes curves that are a superset of those used to diagnose singularities according to the Hawking-Penrose theorem. In the Hawking-Penrose theorem, singularities are diagnosed by incomplete timelike or null geodesics, namely, `freely falling' timelike or null paths that stop short. However, as exemplified by~\citet{geroch_68}, geodesic completeness (understood either as timelike, null, or spacelike geodesic completeness) does not seem to be sufficient for the diagnosis of a singularity: there exist spacetimes that are geodesically complete, but where there remain timelike curves of {\it bounded acceleration} with finite length. [See also~\citet{beem_76} for two further examples of such scenarios.]  Such curves---traversable by an idealized physical observer with a rocket (for example)---seem to be problematic, and the consideration of such curves has indeed guided work on the nature of singularities in particular metric theories of gravity. [See, for example,~\citet{olmo+al_18}.] The Ellis-Schmidt classification scheme covers such a case by diagnosing singularities by curves that satisfy a form of incompleteness that is more general than both geodesic incompleteness and `bounded-acceleration incompleteness', namely, via curves that are $b$-incomplete. The condition of $b$-completeness implies both geodesic completeness and bounded-acceleration completeness.\footnote{A (half) curve is $b$-complete just in case it has infinite generalized affine length. [See~\citet{hawking+ellis_73} for some background.]}

In this paper, we will assume that singularities are indeed diagnosed by (half) curves that are $b$-incomplete---both to avail ourselves of the classification scheme described by Ellis and Schmidt, as well as for the physical character of this definition. Note that, however, when we connect our discussion of singularities to the issue of indeterminism, the sense of the incompleteness of curves that arises---whether it be, for example, geodesic incompleteness, bounded-acceleration incompleteness, or $b$-incompleteness---will be less important than the fact that curves, indeed, stop short.\footnote{Note that the definition of a singularity adopted in the main text does include spacelike $b$-incomplete curves; though there is a question about whether spacelike $b$-incompleteness should be included in one's definition of a singularity. \citet[p.~260]{hawking+ellis_73} define a ``space-time to be  {\it singularity-free} if it is b-complete''; and recognize that one may wish to relax this definition so that it refers to {\it non-spacelike} $b$-completeness. (Ultimately they do not endorse this latter option.)}

\subsection{Causal properties of singularities}\label{SEC:SingularitiesCAU}

Singularities may also be described as either spacelike, timelike, or null~\cite{penrose_74}. We refer to these properties as `causal' as they can be informatively described through the behavior of timelike or null curves (that is, `causal' curves). 

\begin{figure*}
\hspace{-2cm}
\begin{minipage}{.33\linewidth}
\centering
\includegraphics[scale=0.32]{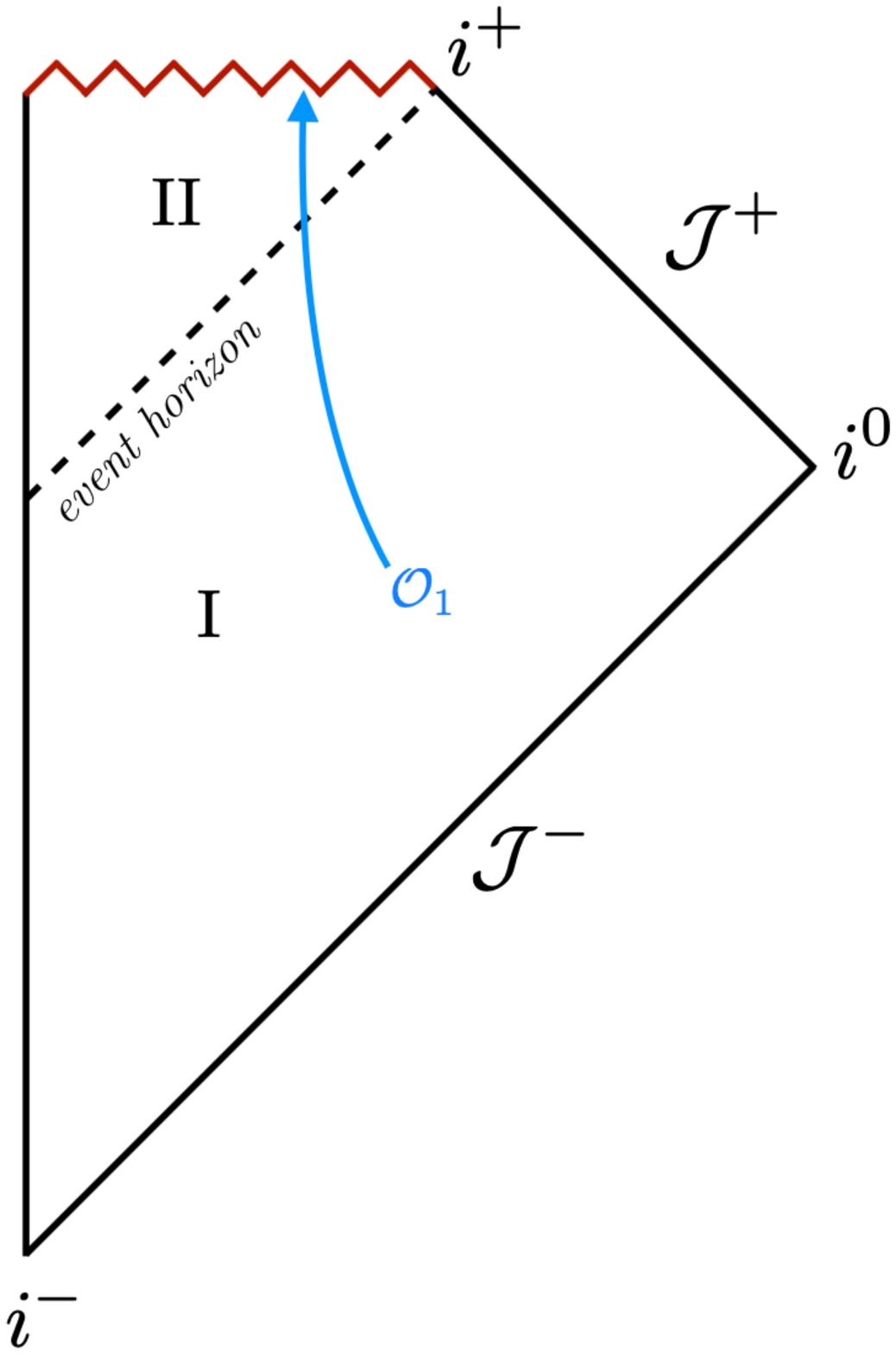}
\\ \vspace{-1cm}(a)
\end{minipage}
\begin{minipage}{.33\linewidth}
\centering
\includegraphics[scale=0.32]{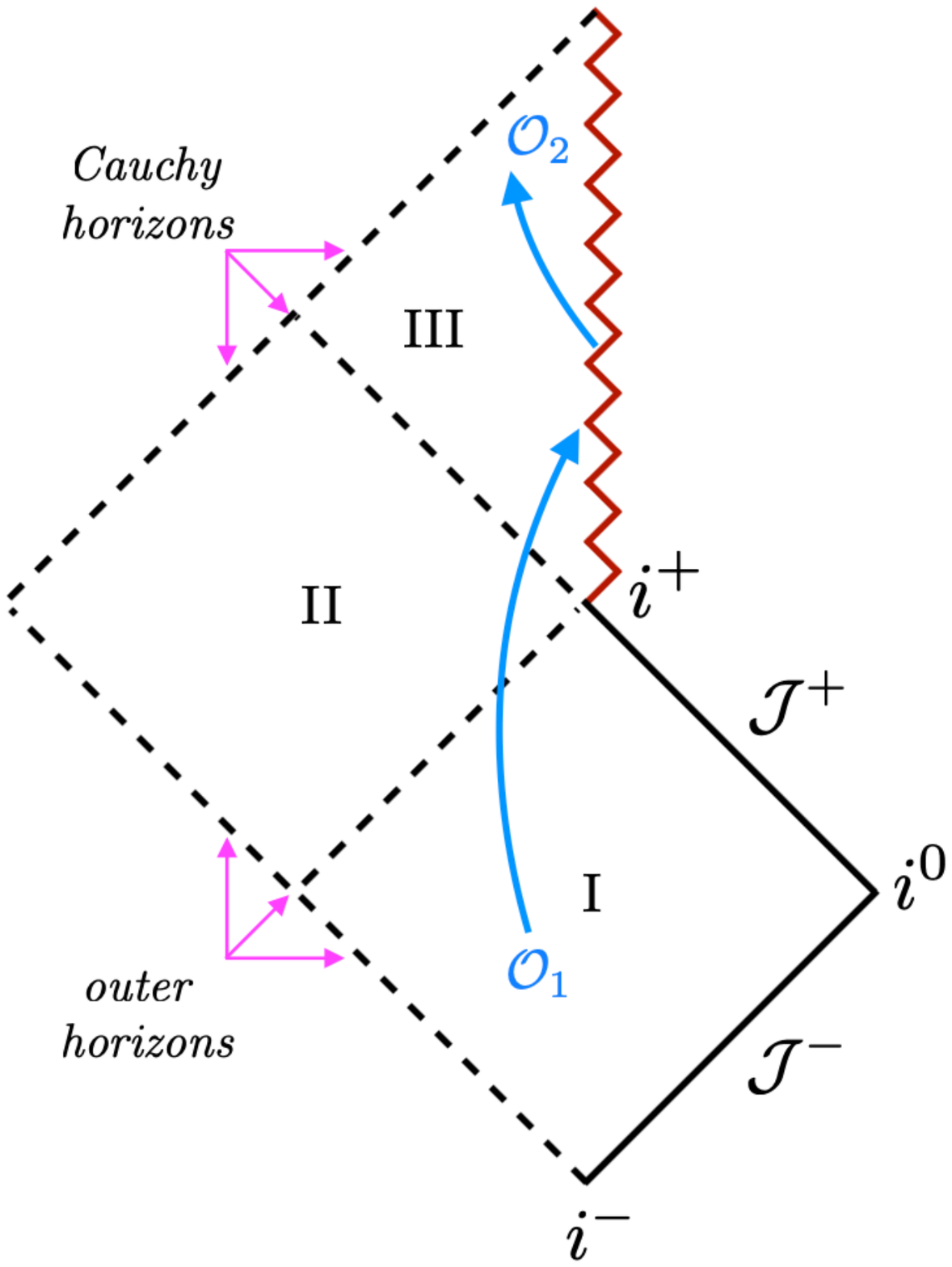}
\\ \vspace{-1cm}(b)
\end{minipage}
\begin{minipage}{.33\linewidth}
\centering
\includegraphics[scale=0.32]{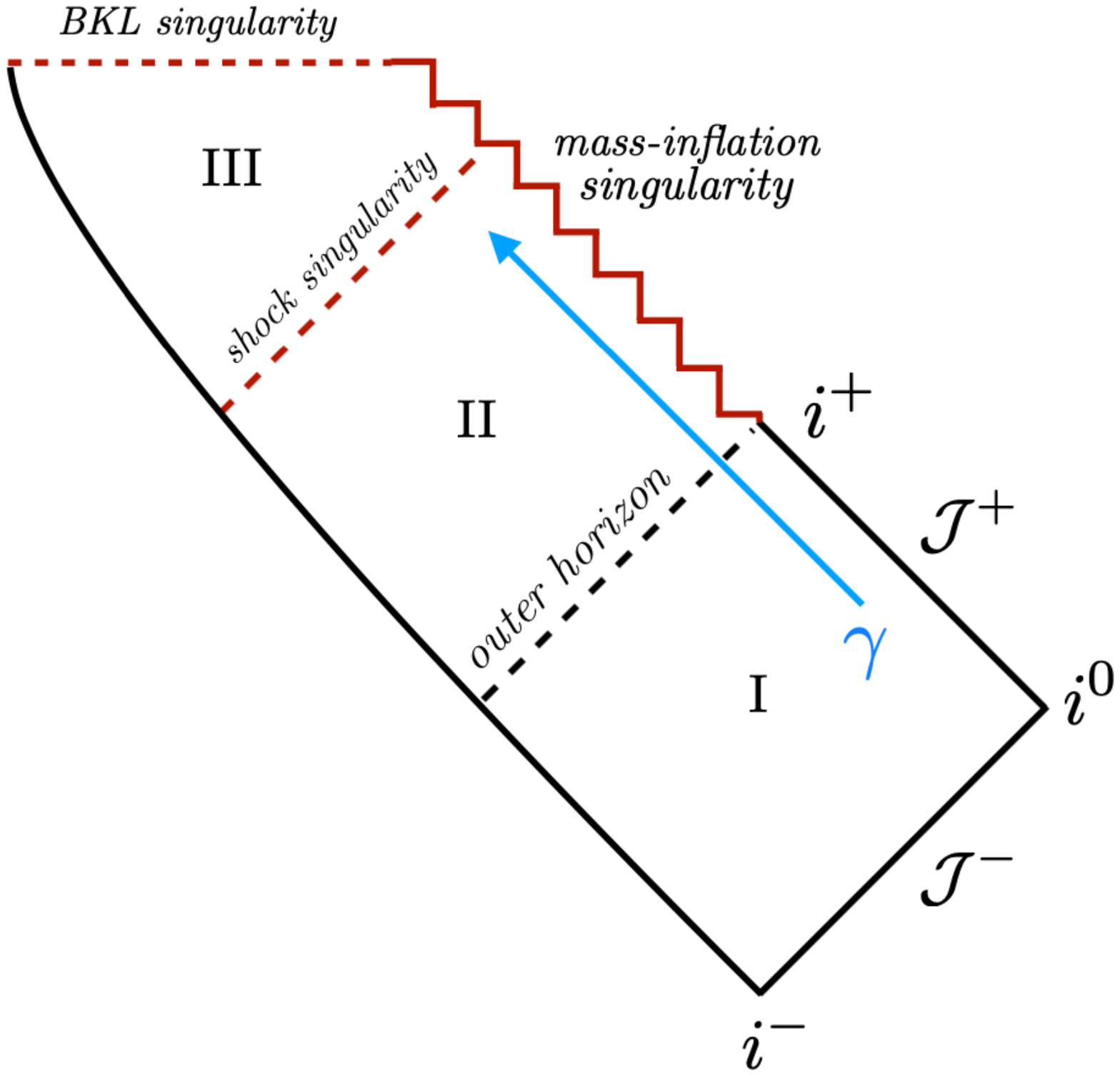}
\\ \vspace{-1cm}(c)
\end{minipage}
\caption{Penrose diagrams illustrating three different types of singularities as they arise in three different types of black holes. In all three cases, regions denoted by I are outside the black hole, whereas regions denoted by II and III are inside the event/outer horizon of the black hole. The red jagged lines denote singularities. Also, for all three diagrams, $i^{-}, i^{0}$, and $i^{+}$ respectively denote past-timelike infinity, spacelike infinity, and future-timelike infinity; whereas $\mathcal{J}^{-}$ and $\mathcal{J}^{+}$ respectively denote past-null infinity and future-null infinity. (a) This diagram depicts the spacetime of a black hole that forms from gravitational collapse and illustrates a spacelike singularity. This singularity serves as an endpoint for the timelike curve $\mathcal{O}_1$. (b) This diagram depicts a portion of the Penrose diagram of a charged black hole (viz.~a Reissner-Nordstr\"{o}m black hole) and illustrates a timelike singularity. This singularity can serve as the past endpoint of a timelike curve (as for $\mathcal{O}_2$) or as the future endpoint of a timelike curve (as for $\mathcal{O}_1$). (c) This diagram depicts an old rotating black hole [adapted from~\citet{scheel+thorne_14}] and illustrates a mass-inflation singularity, which is a null singularity. A photon (denoted by $\gamma$) can pass by the singularity without falling into it.}
\label{FIG:Penrose}
\end{figure*}

Singularities that serve as end-points for timelike curves that would otherwise be future complete are known as future spacelike singularities. (A similar definition holds for past spacelike singularities.) An example of a future spacelike singularity is the singularity that is thought to arise at the center of a black hole formed from the gravitational collapse of matter: a timelike curve that enters the event horizon of such a black hole will eventually `hit' the singularity and will come to an end [see Fig.~\ref{FIG:Penrose}(a)]. Timelike singularities are those such that a timelike curve (as traversed, for example, by an idealized observer) may pass along-side them without falling into them. They can serve as {\it both} the past and future end-points of timelike curves. Such singularities can be found, for example, deep inside a Reissner-Nordstr\"{o}m black hole, as depicted in Fig.~\ref{FIG:Penrose}(b). Finally, null singularities are singularities for which photons (for example) may travel along-side them without falling into them. Such singularities might be found inside the interior of an old black hole---the `mass-inflation singularity' of a Kerr black hole, depicted in Fig.~\ref{FIG:Penrose}(c), corresponds to one such example. 
\\

Thus singularities have geometrical properties (determined by how geometrical quantities, such as the Riemann tensor, behave in their vicinity) as well as causal properties (determined by their relation to causal curves, in their vicinity). In what follows, we will relate our formulation of indeterminism to aspects of these properties.

\section{A novel formulation of indeterminism}\label{SEC:Indeterminism}

In this section we build the main thrust of our argument. In particular, we establish a notion of indeterminism---broader than the standard Laplacian notion---that describes indeterminism as it arises in physical theories, considered generally. We then show that it is this broader notion that comprehensively captures the senses in which indeterminism arises as a result of singularities in GR. As we alluded to in Sec.~\ref{SEC:Intro}, this limitation of the scope of Laplacian indeterminism as it applies to singularities in GR is not necessarily surprising, for such singularities are not usually associated with indeterminism.~\citet[p.~188]{earman_86} presents a reason for why this association is not more prevalent:
\begin{quote}
\dots singularities are an ugly stain on the success of determinism in general relativity. Focus on the subclass of models with Cauchy surfaces. Then by our definition of determinism and the results of the gravitational initial value problem, Laplacian determinism holds. But for models with singularities the victory of determinism has a Pyrrhic flavor, for at best the prediction of singularities is a prediction of the breakdown of the laws of the theory. That breakdown is not countenanced as a breakdown in determinism since the `places' where the singularities occur are not countenanced as part of the arena where determinism wins or loses. The ever more clever means by which determinism avoids falsification are as impressive as its straightforward successes. 
\end{quote}
We argue that the suspect nature of determinism's ``victory'' (alluded to in the above quote) betrays a different state of affairs: namely, there really was no such victory in the first place---indeed, the types of breakdown associated with singularities should also be countenanced as a breakdown in determinism.

\subsection{Indeterminism \`{a} la Laplace}\label{SEC:LD+}

The standard account of determinism can be traced back to~\citet{laplace_1814}. We will describe this account as it relates to theories---as opposed to, say, some external reality directly [see, for example,~\citet{butterfield_05}, who discusses this distinction]. On this account, a theory $\mathcal{T}$ is deterministic iff given a `state', $\mathcal{S}(t)$, of a `physical system', $\mathcal{P}$, at some moment in time, $t$, the future (and possibly the past) states of $\mathcal{P}$ are {\it uniquely prescribed}. Here, $\mathcal{P}$ refers to the {\it entire} `time-evolved' history of $\mathcal{S}$ as understood in the context of the theory $\mathcal{T}$.

This broad definition can be applied to GR under the following identifications.
\begin{itemize}
\item[(i)] $\mathcal{T}$ is GR (in four spacetime dimensions, say).
\item[(ii)] The physical system, $\mathcal{P}$, consists (in principle) of three items: $\mathcal{P}=\langle {M}, g, T\rangle$, where ${M}$ is a four-dimensional manifold, $g$ is a metric tensor, and $T$ is a stress-energy tensor.
\item[(iii)] A state of this physical system, $\mathcal{S}(t)$, at some moment in time, $t$, consists of two items: $\mathcal{S}(t)=\langle \Sigma_{t}, D_{t}\rangle$, where $\Sigma_{t}$ is a (three-dimensional) hypersurface (that is, for example, compact or else asymptotically flat) in $M$, and $D_{t}$ corresponds to suitable initial data. These initial data must satisfy constraint equations (a subset of the full Einstein field equations). They correspond to a three-dimensional metric on $\Sigma_{t}$ and derivatives (in effect, with respect to time) of the metric, as well as matter fields on $\Sigma_{t}$ and corresponding derivatives. 
\end{itemize}
Laplacian determinism then holds for GR when given a state $\mathcal{S}(t)=\langle\Sigma_{t}, D_{t}\rangle$, at $t$, all states $\mathcal{S}(t^{\prime})$ for $t^{\prime} > t$ (and possibly also for $t^{\prime} < t$) are uniquely prescribed. 

Note that states at times other than $t$ are generated by evolving the initial data according to Einstein's equations, so that this characterization of determinism can be understood as a statement about the uniqueness of solutions of the appropriate differential equations. Furthermore, there is a redundancy in GR such that the union of all states for which $t^{\prime} > t$ [namely, the four-dimensional spacetime that results from evolving $\mathcal{S}(t)$] can be related via diffeomorphisms to other observationally indistinguishable  four-dimensional spacetimes. The notion of Laplacian determinism in which we are interested demands uniqueness only up to diffeomorphisms.\footnote{\label{fn:covariant}Thus, we will set aside the type of indeterminism that GR is arguably subject to as a result of the `hole argument'. [For a recent historical account see, for example, Secs. 1 and 2 of~\citet{roberts+weatherall_20}.]}

There are also related versions of Laplacian determinism, which address various choices that have been made in the characterization above. One such version is provided by~\citet{butterfield_89} [see also~\citet{doboszewski_19}] where a broader notion of a `state' is introduced: in particular, the hypersurface $\Sigma_t$ is replaced by a `spacetime region'---which does not have to correspond to a `time-slice'. Indeterminism arises when two ``models'' (where a model consists of a manifold and associated geometrical structures) that agree (in a certain sense) on such a spacetime region, cannot be identified with each other on the entire manifold. [See~\citet[p.~9]{butterfield_89} for further details.] 

In sum, more standard ways of describing indeterminism (that is, Laplacian indeterminism) refer, in essence, to the {\it non-uniqueness} of states of a physical system---despite one having specified another (sufficiently detailed and, for example, antecedent) state. In the following subsection, we will redefine indeterminism, taking a broader point of view.

\subsection{A broader understanding of indeterminism}\label{SEC:Intuit}

Let us step back for a moment and consider, in a general physical setting, a less precise description of determinism [as compared to, for example, the version by~\citet{butterfield_89}, mentioned above]. In particular, consider the following: 
\begin{quote}
Determinism requires a world that (a) has a well-defined state or description, at any given time, and (b) laws of nature that are true at all places and times \dots if (a) and (b) together {\it logically entail} the state of the world at all other times (or, at least, all times later than that given in (a)), the world is deterministic. Logical entailment, in a sense broad enough to encompass mathematical consequence, is the modality behind the determination in ``determinism''~\citep[Sec.~2.5]{hoefer_16}.
\end{quote}
On this account, an indeterministic world is one in which if you specify its state at some time then the (true) laws of nature fail to determine the state of the world at some other time(s). Our conception of indeterminism takes this insight as a starting point. In fact, we need to amend this insight in two ways so that it can apply to the types of cases of interest in this paper. First, we are interested in indeterminism as it applies specifically to theories (as opposed to the {\it world}). Secondly, the above characterization of indeterminism crucially involves a lack of a logical entailment of states at some other time(s). We understand such a lack of entailment (namely, a lack of determination) of a state, in the context of some physical theory, as a {\it failure to `specify' a state}---where this failure can arise in one of three ways. That is, there are three ways in which indeterminism can arise for some physical theory. For ease of later reference, we will denote each such `source' of indeterminism by a mnemonic label.
\begin{itemize}
\item[] {\textsc{Non-uniqueness}}: 
\begin{quote}The theory provides multiple, distinct accounts of future (or past) states of a physical system.
\end{quote} 
\item[]{\textsc{Non-existence}}: 
\begin{quote} The theory provides {\it no account} of some possible future (or past) state of a physical system.
\end{quote} 
\item[]{\textsc{Incoherence}}: 
\begin{quote}The theory does provide an account of future (or past) states of a physical system, but that account fails to satisfy conditions (viz.~desiderata), $\mathcal{C}$, that are internal and/or external to the theory.

The criteria $\mathcal{C}$ provide a check on whether the specification of a state by the theory leads to: (i) an internal conflict, in which case we refer to the incoherence as {\textsc{Internal incoherence}}; and/or (ii) a conflict with some pre-established theory (or principle) that enjoys broad support---a case we dub {\textsc{External incoherence}}.
\end{quote} 
\end{itemize}
Note that Laplacian determinism only captures \textsc{Non-uniqueness}. The other two sources are thus responsible for the claim that our formulation is broader than the usual Laplacian one. We now turn to some salient aspects of the above characterization of indeterminism.

As regards {\textsc{Non-uniqueness}}, the sense in which future (or past) states are `distinct' can, of course, be understood in the context of the particular theory (though possibly in multiple, prima facie different, ways). For example, in Newtonian dynamics, {\textsc{Non-uniqueness}} can arise as a result of the time-reverse of a scenario [described by~\citet{xia_92}] in which a particular configuration of five point-masses moving in three-dimensional Euclidean space can escape to spatial infinity in a finite time. Here, distinct accounts of future states arise depending on whether or not particles do, indeed, fly in from spatial infinity. [See, for example,~\citet{hoefer_16} for a description of such `space-invaders' and~\citet{werndl_16} who describes the resulting lack of uniqueness in this way.] In the case of GR, one can diagnose {\textsc{Non-uniqueness}} through, for example, the lack of a Cauchy surface for a spacetime manifold. The sense in which accounts of future states are distinct can then be understood in terms of the existence of non-diffeomorphic specifications of spacetime regions beyond the Cauchy horizon.

Perhaps the most contentious assertion we have made is due to {\textsc{Non-existence}}, wherein we claim that it is not consistent with determinism for a theory to leave the description of some system unspecified. Previewing our discussion in Sec.~\ref{SEC:Time}, such a situation arises in GR for singularities, when end-states of particles traveling along curves are left unspecified---such as when one traces such curves back into the big-bang singularity. That this signals indeterminism breaks with the common intuition to which Earman refers, quoted in the preamble to this section, wherein `places' where singularities arise aren't included ``as part of the arena where determinism wins or loses''. In contrast, we assert that those places where singularities `arise' should be included as part of the arena where determinism wins or loses: leaving unspecified the end state of a particle moving along an incomplete trajectory is a key (definitional) aspect of a {\it lack of determination}.\footnote{\label{fn:DetArena}Note that we are not suggesting that singularities should necessarily be considered to be part of spacetime---that is, we are not necessarily enlarging our conception of spacetime to include singularities; we are advocating enlarging the arena over which determinism wins or loses.} (We will elaborate on this issue in Sec.~\ref{SEC:Time}.)

As regards \textsc{Internal incoherence}, it has been argued by~\citet{vickers_13} that `internal inconsistency'
is difficult to diagnose.\footnote{We have chosen to generally use the term `incoherent' instead of `inconsistent' as we have in mind a notion that is broader than something akin to logical contradiction (which is readily brought to mind by the latter term).} We will not need to take a stance on this claim---but point out only that {\it if} the specification of a state by a theory is internally incoherent (or, indeed, internally inconsistent) then that should count as a failure of the theory to {\it specify} the state. Nevertheless, there are clear cases where {\textsc{Internal incoherence}} can arise: as in certain solutions of GR that contain closed timelike curves. Here, the possibility arises of an explicit contradiction (that is indeed internal to the theory): namely, of the existence of some state and the nonexistence of that same state (as for the usual paradoxes that relate to time-travel, such as the `grandfather paradox').\footnote{See~\citet{sklar_90}, for an interesting discussion of the impact, if any, of such paradoxes on the (initial) conditions that may give rise to such a situation.} Two examples of {\textsc{External incoherence}}---with differing fates for the proposed model---are: (a) Bohr's model of the atom, which when proposed was in conflict with Maxwellian electrodynamics~\cite{vickers_13}; and (b) the steady-state model of the universe, with its need for the creation of matter (so that the density of the expanding universe would remain constant), which was in conflict with the principle of conservation of energy~\cite{mcmullin_82}. As we will elaborate on in the following section---both senses of {\textsc{Incoherence}} will arise in describing singularities in GR.

Finally, note that our formulation of indeterminism can be used to assess indeterministic features of individual models themselves (where, by a `model', we mean a specific solution of the equations that one derives from the theory of interest). So, for example, in the context of GR and in metric theories of gravity more generally, {\textsc{Non-existence}} and {\textsc{External incoherence}} can be applied to specific manifolds and the fields defined on these manifolds, to diagnose the existence of indeterminism in these models. For example---and previewing a claim that will arise in the following section---a specific set of initial conditions that yields a region in which the end-state of a particle traveling along a timelike curve is unspecified reveals a specific model that is indeterministic, by virtue of it satisfying {\textsc{Non-existence}}. For cases that arise exclusively as a result of {\textsc{Non-uniqueness}}, at least two distinct models are required to diagnose indeterminism, though, of course, one can label a single spacetime as `non-unique'.

We now further exemplify and apply our formulation  of indeterminism specifically to features of singularities as they arise in GR.

\section{Indeterministic features of singularities in general relativity}\label{SEC:Sing}

There are a variety of features of singularities in GR that, we contend, exhibit indeterminism, in the style of that described in Sec.~\ref{SEC:Intuit}.\footnote{See~\citet{doboszewski_19} for another account of  indeterministic features of GR more generally.} In this section we will provide an account of such features, including a discussion of why they indeed signal indeterminism. Furthermore, we will  highlight how this indeterminism relates to the Laplacian notion discussed in Sec.~\ref{SEC:LD+}. 

\subsection{Unspecified end-states}\label{SEC:Time} 

As described in Sec.~\ref{SEC:SingularitiesGEO}, the {\it defining property} of singularities, as they are commonly understood, is that they bring curves to an end---viz.~such curves cannot be extended indefinitely: they are {\it incomplete}. These curves fall into one of three classes: timelike, null, or spacelike. For singularities that are diagnosed via the former two classes (that is, via {\it causal} curves) one encounters  future and/or past incompleteness. Consider the case in which a singularity is diagnosed by a timelike or null future-incomplete curve. This can be realized by a test particle that travels along the curve and eventually encounters the singularity.\footnote{For a timelike future-incomplete curve, the test particle can be understood, for example, as an idealized observer. More generally, the particle can be described as part of the evolving state of the physical system, where it does not back-react on the underlying geometry and where it has a sufficiently small size (so that one can describe its trajectory via a curve).} Importantly, a consequence of the incompleteness of the curve is thus that, according to the theory, the {\it end-state} of a particle moving along such a trajectory is left unspecified. 
Therefore, we contend, indeterminism arises as a result of {\textsc{Non-existence}}. 
 
One may agree with a claim in the quote of  Earman's (as discussed at the outset of Sec.~\ref{SEC:Indeterminism}) that there is no `place' (or `time') for which GR needs to provide an account of the particle, when that particle's worldline indeed comes to an end in a singularity; yet we maintain that the scenario manifests indeterminism (see fn.~\ref{fn:DetArena}). There are questions one can naturally ask for which the theory does not provide an answer. More explicitly (but rather colloquially), one can ask about the fate of the particle---which, prior to encountering the singularity,  possessed a well-defined state. For example, does the particle indeed have a finite lifetime? If so, what does it mean, according to the theory, for the particle to not exist: that is, is there some state describing this {\it non-existence} in the space of possible states? If not, then what happens to the particle? Moreover, one may wonder about the process of `hitting the singularity'.\footnote{See, for example,~\citet[p.~152]{hawking+ellis_73},~\citet[p.~193]{wald_84}, and~\citet[p.~227]{carroll_GR}, where such language is used or implied. (We do not, of course, assume the authors imply, by the use of such language, the existence of some GR-based account of what the process of `hitting the singularity' entails.)} To reiterate: the theory does not provide an answer to such questions.
 
Note that the situation here is very different from what happens, for example, in quantum field theory (QFT). In the latter context it is common to consider scenarios in which particles are annihilated (or are created) via a suitable interaction. In this sense, the particle's worldline may ``stop short"---but in contrast with what happens in the case of incomplete curves in GR, the  annihilation and creation of particles can be described within QFT. In fact, the state of the ``physical system" remains well-defined even after the disappearance of a particle.

Analogous pathologies may arise in the case of past incompleteness (which, one might argue, are more problematic). For example, particles may enter spacetime from a singularity and interact with other particles~\citep{penrose_69, shapiro+teukolsky_91, earman_95}. This is a general-relativistic version of the `space invaders' scenario, described in Sec.~\ref{SEC:Intuit}, which arguably provides a source for indeterminism in Newtonian mechanics. Indeterminism arises because GR does not provide an account of the {\it process or mechanism} by which such matter may form at/near singularities; thereby, again, satisfying conditions for {\textsc{Non-existence}}.\footnote{Note that scenarios resulting from {\it whether or not} particles are emitted by singularities, could be understood in terms of \textsc{Non-uniqueness} (in much the same way that scenarios resulting from `space invaders' in Newtonian dynamics, discussed in Sec.~\ref{SEC:Intuit}, are usually understood). Here (that is, in the main text above) we have attributed the indeterminism that arises to \textsc{Non-existence} as, we contend, this source is ultimately responsible for the indeterminism.}

How is this {\textsc{Non-existence}} related to a Laplacian notion of determinism? Indeed, it is not that the end-state of the particle is specified in ways that are non-unique. Rather, the end-state is not specified at all. Laplacian determinism is not compromised by the existence of such incomplete curves. [See also~\citet{doboszewski_19}.]

\subsection{Ill-behaved physical quantities}
\label{SEC:PhysQuant}

`Curvature singularities' in GR, described in Sec.~\ref{SEC:SingularitiesGEO}, are associated with the phenomenon that, in their vicinity, certain physical quantities (i) oscillate without limit or (ii) grow without bound. (Note that this is not true of singularities categorized, in Sec.~\ref{SEC:SingularitiesGEO}, as `quasi-regular singularities'.) 

With regard to (i), for a version of Taub-NUT spacetime, scalar quantities may remain bounded along some curve as one approaches the singularity, but their values oscillate without approaching a limit [see, for further details,~\citet[pp.~38--40]{earman_95}]. In such cases, differentiability constraints (such as the assumption that the underlying manifold is smooth, viz.~$C^{\infty}$) can be violated. The sense of indeterminism that arises is different to that described in Sec.~\ref{SEC:Time}; namely, it is {\textsc{Internal incoherence}} that arises.

With regard to (ii), perhaps the most well-known example is at the center of the simplest type of black hole (a non-spinning, non-charged black hole) in which a quantity that characterizes properties of the curvature of the spacetime in a coordinate invariant way, namely, the Kretschmann scalar ($R_{\mu\nu\sigma\tau} R^{\mu\nu\sigma\tau}$) diverges as $1/r^{6}$ (here, $r$ measures radial distance from the `center' of the black hole).\footnote{We make no claims about the  severity of the pathologies outlined in Sec.~\ref{SEC:Time} compared with those in Sec.~\ref{SEC:PhysQuant}, though it is interesting to note the following claim, that relates to such a comparison: ``Timelike geodesic incompleteness has an immediate physical significance in that it presents the possibility that there could be freely moving observers or particles whose histories did not exist after (or before) a finite interval of proper time. This would appear to be an even more objectionable feature than infinite curvature and so it seems appropriate to regard such a space as singular''~\citep[p.~258]{hawking+ellis_73}.} The sense of indeterminism at play in such a scenario is again different to that described in Sec.~\ref{SEC:Time}. In particular, at some point (for example, at some radial distance $r_{*}$ from the `center' of the  black hole---or for certain sufficiently large values of a parameter that describes points along a curve headed into the singularity), the values taken by physical quantities are not to be taken literally: the theory supplies values for physical quantities that are `too large' according to certain criteria that we demand of GR (that are not encoded in the theory itself). The theory is subject to a charge of {\textsc{External incoherence}}. Note that this charge of indeterminism is leveled at GR despite the fact that the theory can (and does) specify a unique value for quantities such as invariant scalars---again (as in Sec.~\ref{SEC:Time}) a sense of Laplacian determinism is not being compromised.

Of course, one may wonder about the external standard(s) by which we judge whether certain physical quantities should be taken literally. There are two points of view, one more phenomenological, the other more formal. In the former case, the claim is that such divergences do not appear to arise in nature and so when a theory predicts that they do, something has gone wrong. The more formal response can be understood in one of two ways.
\begin{itemize}
\item[(i)] First, dimensional analysis can reveal where quantum gravity effects are expected to be important (that is, where the classical theory is expected to break down)---in particular, when values for the curvature approach those associated with the Planck scale [for example, $|R| \sim 1/l_{\rm Pl}^2 \equiv c^3/(\hbar G)$, where $l_{\rm Pl}$ denotes the Planck length]. Such an analysis thus provides external conditions that can diagnose {\textsc{External incoherence}}. 
\item[(ii)] A second response (that is, in principle, different from the first) is related to how we can conceive of GR as the leading-order contribution to a more general effective theory. In certain situations, the effects of higher-order contributions can become as important as those of GR: the effective theory expansion thus breaks down and GR is no longer descriptively accurate as the leading-order contribution. Again, this situation manifests {\textsc{External incoherence}}.

In a little more detail, for such effective-field-theoretic extensions of GR, one typically writes down an action by identifying the relevant symmetries and ordering the action according to terms that contribute sequentially smaller amounts to the action at lower energies (in particular, below some cutoff in the energy scale). For pure gravity, the symmetries are general coordinate invariance, namely, invariance under four-dimensional spacetime diffeomorphisms. One can extend GR via an action of the following form [see, for example,~\citet{donoghue_94} for further details]:
\begin{equation}\label{EQN:EFTAction}
S = \int d^{4}x \sqrt{-g}\left(\Lambda + c_0 R + c_1 R^2 + c_2 R_{\mu\nu}R^{\mu\nu}+\dots \right),
\end{equation}
where $\{\Lambda,\{c_i\}\}$ are constants to be determined, for example, from experiment. (We have left out a Lagrangian for matter degrees of freedom as that will not be important in what follows.) The four terms displayed in parentheses in Eq.~(\ref{EQN:EFTAction}) have, respectively, zero, two, four, and four derivatives of spacetime coordinates and so---as is usual in EFT-energy expansions---these last two terms are less important for describing physics below the energy cutoff of the EFT expansion (which is often taken to be the Planck energy-density---but, indeed, does not have to be). Such an expansion provides conditions in which GR  might be deemed to be externally incoherent (that is, it would be deemed to be incoherent with the above-defined effective-field-theoretic generalization). For example, if when traversing some curve in a manifold that comprises a solution to the Einstein field equations, the second term in the action above becomes comparable to either of the next largest terms, viz.~$|c_0 R|\sim {\rm min}\{|c_1 R^2|, |c_2 R_{\mu\nu}R^{\mu\nu}|\}$, then the perturbative expansion breaks down---and the claimed (external) incoherence arises.
\end{itemize} 

\subsection{Lawlessness}\label{SEC:Forbid}

GR is effectively `silent' about occurrences at (or sufficiently close to) singularities, so there are, as noted by~\citet{penrose_69}, no `laws' of singularities. One cannot study singularities within GR as entities in and of themselves, in the same way that one might use the standard model of particle physics to probe properties of yet undiscovered particles. As a consequence, a form of indeterminism arises. For example, GR is effectively silent about the possibility of matter/energy emanating from singular regions (a process not expressly forbidden by GR---but about which GR does not make any positive claims).\footnote{In this way, singularities can be (at least) locally visible, namely, they can be {\it locally naked}. (Note that it is possible, in principle, for such singularities to arise behind an event horizon.) The {\it weak cosmic censorship conjecture} (WCCC), formulated by~\citet{penrose_69}, provides a well-known (potential) limitation on how visible certain singularities may be (in the context of GR). Containing the scope of the resulting putative indeterminism is the primary goal of the WCCC. Despite a large (and growing) literature, the WCCC remains---50 or so years after it was first described---an open problem. [See the following select set of references:~\citet{geroch+horowitz_79, earman_95, wald_97, christodoulou_99, hod_08, joshi+malafarina_11}.]} This renders undetermined other regions of spacetime in causal contact with singular regions. A second perhaps more exotic possibility also presents itself. Namely, GR does not account for the possibility of the evolution of the nature of singularities themselves. That is, the theory is also silent about whether singularities can evolve in such a way as to change their geometric and/or causal properties\footnote{As we will discuss in Sec.\ref{SEC:Varying}, certain physical arguments suggest that such evolution is indeed possible, though GR does not provide a detailed account of the actual process.}; or whether singularities can cease to exist, giving rise to non-singular spacetimes.\footnote{As noted at the outset, we are focusing here on classical considerations. Taking quantum effects into account may result in the evaporation of black holes (as well as their singularities) via the emission of Hawking radiation. However, it is not clear that the emission of Hawking radiation persists when the mass of the black hole becomes comparable to the Planck mass, and thus whether black holes and their singularities evaporate completely. (Here, the semi-classical approximation is expected to break down and quantum gravitational effects are expected to become important.) So, despite taking into account such considerations that generalize GR, we do not have a definite answer to the question of whether singularities can cease to exist.} In this way, the theory satisfies conditions for {\textsc{Non-existence}}---in that it stops providing a description of physical regions near and moreover `at' singularities.

\subsection{Cauchy horizons}\label{SEC:CH}

The final source of indeterminism we will mention relates to Cauchy horizons in GR [as depicted in, for example, Fig.~\ref{FIG:Penrose}(b)]. Generally, these correspond to boundaries of spacetime beyond which the initial value formulation of GR (alluded to in Sec.~\ref{SEC:LD+}) does not uniquely specify the state of spacetime. 

Certain types of singularities lead to Cauchy horizons. In particular, they can arise when one has a timelike singularity or a null singularity. As far as we are aware, spacelike singularities only arise at the boundaries of spacetime---in such a way that no Cauchy horizon arises [see~\citet{penrose_74}]. (So, for example, there is no Cauchy horizon associated with a big-bang-type singularity.) The sense of indeterminism that arises corresponds to a clear violation of determinism understood in a Laplacian sense: such a scenario satisfies conditions for {\textsc{Non-uniqueness}}.\\

In sum, as regards the four indeterministic features of singularities described above, three of the features have aspects that are not consistent with a Laplacian notion of indeterminism (see Table~\ref{TAB:Summary}). Perhaps most notably, Laplacian indeterminism fails to capture what is arguably a key feature of singularities, namely, that they stop particles in their tracks in such a way as to leave the end-state of the particle undetermined. Laplacian determinism does not therefore provide a general way to describe indeterministic features of singularities in GR. [Note that this assessment is consistent with~\citet{doboszewski_19} who takes issue with Laplacian determinism as a way to provide a general account of determinism in GR.]\footnote{The question of the existence of singularities as they may arise in more general theories of gravity, in particular for general metric theories of gravity, is an interesting one. Specific analyses do exist, and these analyses diagnose singularities via incomplete curves. According to our formulation of indeterminism, therefore, such settings also signal indeterminism. [See, for example:~\citet{horowitz+myers_95, alani+santillan_16, bejarano+al_17}.] We leave a more detailed analysis of these issues for future work.}

\begin{table}
\begin{tabular}{l @{\hskip 12pt} c@{\hskip 10pt} c@{\hskip 10pt} c}
&\multicolumn{3}{c}{\it Sources of indeterminism}\\
\cline{2-4}
{\it Features of singularities}    & {\textsc{Non-uniqueness}} & {\textsc{Non-existence}} & {\textsc{Incoherence}}\\
\hline
{Unspecified end-states} 			& --		& \checkmark    & --	\\
{Ill-behaved physical quantities}    	& --		& --     	 & \checkmark	\\
{Lawlessness}     		& --	& \checkmark	 & --	\\
{Cauchy horizons}				& \checkmark	& -- 		 & --	\\
\end{tabular}
\caption{\label{TAB:Summary} Sources of indeterminism (as described in Sec.~\ref{SEC:Intuit}) for the four features of singularities discussed in the main text (in Secs.~\ref{SEC:Time}--\ref{SEC:CH}). The left-hand column lists these features and the remaining columns catalog whether the particular source of indeterminism is present (\checkmark) or not (--).}
\end{table}

\section{Transitions between sources of indeterminism}\label{SEC:Varying}

An interesting feature of indeterminism, as it manifests due to singularities in GR, is that the source of the indeterminism in a particular physical setting can change.

For instance, dropping a charged particle into a Schwarzschild black hole converts the black hole into that of a Reissner-Nordstr\"{o}m black hole. [So, for example, a Penrose diagram such as that depicted in Fig.~\ref{FIG:Penrose}(a) is transformed into one such as that (partially) depicted in Fig.~\ref{FIG:Penrose}(b).] The central singularity is originally a spacelike singularity in which (i) the end-state of a particle that travels along a curve that heads toward the singularity is not specified and (ii) physical quantities, such as the Kretschmann scalar, are unbounded. Indeterminism arises due to {\textsc{Non-existence}} and {\textsc{External incoherence}}, respectively. After dropping the charged particle into the Schwarzschild black hole, this singularity is converted into a timelike singularity (with an associated Cauchy horizon) in which indeterminism arises---in addition to {\textsc{Non-existence}} and {\textsc{External incoherence}}---as a result of {\textsc{Non-uniqueness}.\footnote{An analysis similar in spirit can be given, for example, for a particle dropped into a Schwarzschild black hole with angular momentum; thereby converting the Schwarzschild black hole into a Kerr black hole.}

Another example arises in considering Cauchy horizons. Note that one can have Cauchy horizons ostensibly without the presence of singularities. But there are interesting cases in which such horizons are, in fact, conjectured to be related to singularities. For example, certain black-hole solutions with Cauchy horizons, such as the Kerr and Reissner-Nordstr\"{o}m solutions, are conjectured to exhibit a `blue-shift instability' of these horizons [see, for example,~\citet{penrose_68} and~\citet{poisson+israel_90}]. Namely, small perturbations in initial data are conjectured to build-up at those horizons to convert them into {\it curvature singularities}.\footnote{~\citet[p.~1797]{poisson+israel_90} phrase a consequence of the satisfaction of this conjecture rather memorably: ``The Cauchy horizon is the ultimate brick wall at which the evolution of spacetime is forced to stop''.} In such an instance, the usual source of indeterminism associated with Cauchy horizons (independent of their relation to singularities), namely, {\textsc{Non-uniqueness}} is, under our formulation of indeterminism, traded for {\textsc{Non-existence}} (due to curves stopping short at curvature singularities where an end-state is unspecified) and {\textsc{External incoherence}} (due to physical quantities being ill-behaved). In short, the presence of singularities that arise as a result of the conjectured blue-shift instability affects the nature of the indeterminism that is associated with the horizon region. This consequence of our approach to determinism invalidates the usual stance on how such an instability impacts the deterministic nature of GR. For example, our approach calls into question the following claim by~\citet{hollands+al_20} (in the abstract of their paper), in which the satisfaction of the strong cosmic censorship hypothesis is associated with avoiding indeterminism:
\begin{quote} The strong cosmic censorship conjecture asserts that the Cauchy horizon does not, in fact, exist in practice because the slightest perturbation (of the metric itself or the matter fields) will become singular there in a sufficiently catastrophic way that solutions cannot be extended beyond the Cauchy horizon. Thus, if strong cosmic censorship holds, the Cauchy horizon will be converted into a `final singularity,' and determinism will hold~\cite[p.~1]{hollands+al_20}.
\end{quote}
Under our formulation of indeterminism, this conversion of the Cauchy horizon into a `final singularity' does not save determinism: it trades one source of indeterminism ({\textsc{Non-uniqueness}}) for another (a combination of {\textsc{Non-existence}} and {\textsc{External incoherence}}).

\section{Envoi}\label{SEC:Conclusion}

For physical theories generally, we have described a notion of indeterminism that takes a broader point of view than the usual Laplacian notion. Under our formulation, as developed in Sec.~\ref{SEC:Intuit}, indeterminism arises when at least one of the following three conditions obtains: (i) a theory does not provide a unique account of states of the physical system under consideration (what we term {\textsc{Non-uniqueness}}); (ii) a theory does not specify a physical state (what we term {\textsc{Non-existence}}); (iii) a theory provides an account of physical states that does not satisfy certain desiderata (what we term {\textsc{Incoherence}}). These three sources provide a more comprehensive characterization---compared to the usual Laplacian notion, which only instantiates \textsc{Non-uniqueness}---of the relevant sense of a lack of `determination' in physical scenarios. 

We have further argued that these sources of indeterminism capture the sense in which a lack of determination arises for a broad set of features associated with singularities in GR. There are four such features, described in Sec.~\ref{SEC:Sing}, namely: (a) unspecified end-states of particles; (b) ill-behaved physical quantities in the vicinity of singularities; (c) a lack of `laws' of singularities; and (d) Cauchy horizons, which arise as a result of certain types of singularities. In Table~\ref{TAB:Summary}, we provide a summary of the associations for which we have argued, between the three sources of indeterminism mentioned above and these four features of singularities. Notably, only for one of these features---namely, for Cauchy horizons---is the source of indeterminism associated with the usual Laplacian notion.

Thus, we conclude that one way to interpret the type of breakdown that is usually attributed to the existence of singularities in GR is, indeed, as a failure of {\it determinism}.

\begin{acknowledgments} We thank Bahram Mashhoon and Mahdiyar Noorbala for discussions. FA acknowledges support from: the Black Hole Initiative at Harvard University, which is funded through a grant from the John Templeton Foundation and the Gordon and Betty Moore Foundation; and the Faculty Research Support Program (FY2019) at the University of Notre Dame.
\end{acknowledgments}

\end{document}